# High temperature oxidation behavior of disordered $(Ti_{0.5}Zr_{0.5})_2AlC$ MAX phase via a Machine Learning-Augmented DFT Approach

P. Singh,[a] D. Sauceda,[a] and R. Arroyave[a,b]

[a]*Department of Materials Science & Engineering, Texas A&M University, College Station, TX 77843, USA*
[b]Department of Mechanical Engineering, Texas A&M University, College Station, TX 77843, USA

**Abstract**

The Zr-based MAX phases have attracted considerable attention for their outstanding irradiation behavior and high neutron transparency relevant to nuclear power generation technologies. In spite of increased understanding of physical behavior crystalline MAX phases, the high-temperature oxidation behavior and reaction mechanism of disordered MAX phases both from theory and experiments are not well understood due to increased system complexity. Here, we present a detailed comparative assessment of high-temperature thermodynamic-stability and oxidation behavior (reaction-products and chemical activity) of ordered $Ti_2AlC$ and disordered $(Ti_{0.5}Zr_{0.5})_2AlC$. We believe that the new insights will enhance our understanding of oxidation process in disordered MAX phases.

---

+**Current Address**:Ames Laboratory, US Department of Energy, Iowa State University, Ames, IA 50011, USA

## Introduction

The MAX phases have received considerable attention due to their unusual properties in harsh surroundings such as exposure to highly reactive chemical environment, high-temperatures, or extreme radiation [**1-3**]. For example, cladding on zirconium-based MAX phases are proposed for improved accidental tolerance in future nuclear systems [**4,5**]. The general formula for MAX phases is $M_{n+1}AX_n$, where the M, A, and X are early transition metals, A group elements, X is carbon/nitrogen, and n=1, 2 or 3 [**6,7**]. While detailed work has been done to understand the oxidation behavior of ordered MAX phase alloys [**8**], the effect of disorder on oxidation mechanism is still a less explored territory both from theory and experiments due to increased complexity of reaction mechanism. Recently, it has been found that compositional disorder can be tuned to achieve desirable properties such as thermoelectricity [**9**], magnetoresistance [**10**], and charge-transport [11]. This suggests that chemical disorder can have an impact on the oxidation behavior of MAX phases.

In this letter, we used a machine learning-based high-throughput scheme to investigate the effect of chemical disorder on oxidation behavior of $(Ti_{0.5}Zr_{0.5})_2AlC$. The Zr-based MAX phase was chosen due to their application as future structural materials [12,13]. The reaction products formed during oxidation process of $(Ti_{0.5}Zr_{0.5})_2AlC$ was analyzed and compared with $Ti_2AlC$ to understand the effect of disorder. Detailed understanding of oxidation mechanism is important for successful design of MAX phases for high temperature application.

## Methods

**Grand Canonical Linear Programming (GCLP):** The OQMD [14] was used as starting point in our high-throughput scheme (details of the framework are presented elsewhere) [15] for Gibbs formation energy ($\Delta G_{form}$) and chemical activity prediction across the temperature range. The GCLP minimizes the free energy of the mixture at a given Ti/Zr/Al/C/O to identify the thermodynamically equilibrium phases [16,17]:

$$\Delta G = \sum_{phase} f_{phase} . \Delta G_{phase}$$

where, $\Delta G_{phase}$ is the free-energy of competing phases, and $f_{phase}$ is the phase-fraction.

**Density-functional theory (DFT):** A 200 atom disorder supercell of $(T_{0.5}iZr_{0.5})2AlC$ was generated using ATAT (see **Fig. S1**) [18]. Full relaxations and charge self-consistency of disorder MAX phase was done using DFT as implemented in Vienna *Ab-initio* Simulation Package [19,20] on 3x3x3 and 5x5x5 Monkhorst-Pack [21] k-mesh, respectively. The PBE exchange-correlation [22] with 533 eV planewave cut-off energy was used in all calculations. Forces and total energies were converged to $-10^{-3}$ eV/Å and $10^{-5}$ eV/cell, respectively. The AFLOW algorithm of convex hull calculation was also utilized [23].

## Results & Discussion

The Gibbs formation energy ($\Delta G_{form}$) is a thermodynamically defined quantity that indicates the *intrinsic* stability of MAX phases—relative to constituent elements—as a function of temperature. The trend in $\Delta G_{form}$ of $(Ti_{0.5}Zr_{0.5})_2AlC$ and their competing phases are shown in **Fig. S2&S3**. The phase and phase-fractions for $(Ti_{0.5}Zr_{0.5})_2AlC$ formed during oxidation reaction are shown in **Fig.**

**1.** The heat map shows competing phases for varying oxygen content from 300 K – 2000 K. All thermodynamically stable binary/ternary phases were included, which were calculated using convex-hull algorithm of OQMD [**14**] and AFLOW [**23**]. The $(Ti_{0.5}Zr_{0.5})_2AlC$ shows to two chemical activity zones, (i) low to moderate oxygen, and (ii) high oxygen, i.e., for low to moderate oxygen, $Al_2O_3$ phase was found at all temperatures, but at high oxygen $Al_2O_3$ decomposes into $TiAl_2O_5$ spinel phase. At the onset of oxidation, $(Ti_{0.5}Zr_{0.5})_2AlC$ decomposes into ZrC, $Ti_3AlC_2$, $Al_2O_3$ and $ZrO_2$. The $Ti_3AlC_2$ phase that disappeared due to high Al oxidation at low oxygen content in $Ti_2AlC$, however, remains stable in disordered $(Ti_{0.5}Zr_{0.5})_2AlC$ as Zr suppresses the Al activity (see **Fig. S4**).

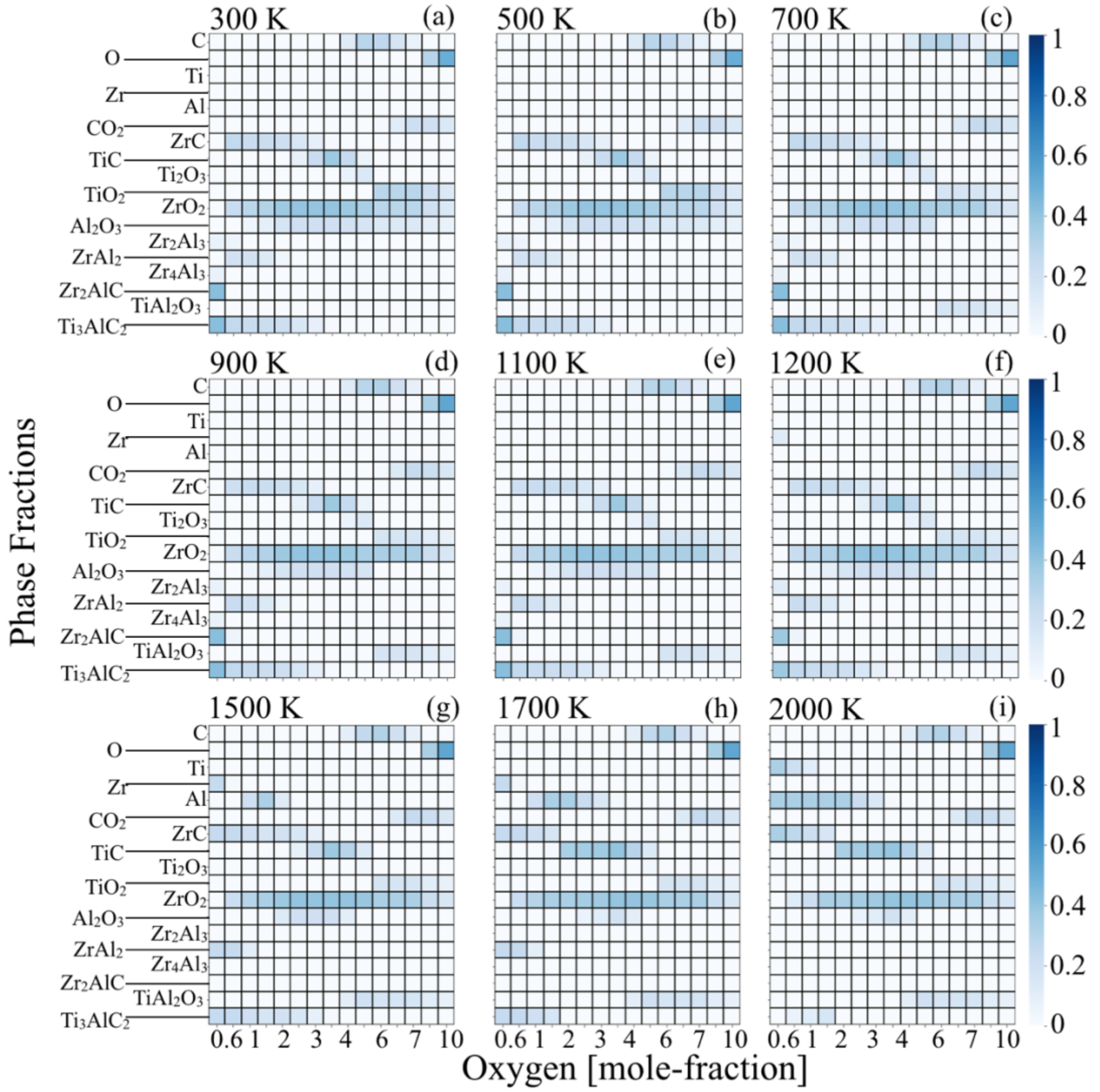

**Figure 1.** (a-i) Phases and phases-fractions during oxidation of $(Ti_{0.5}Zr_{0.5})_2AlC$ from 300 K– 2000 K for varying molar percent oxygen (0-11 moles). The white spot shows missing phase.

The thermodynamically stable reaction products for $Ti_2AlC+O_2$ and $(Ti_{0.5}Zr_{0.5})_2AlC+O_2$ are shown in **Table 1a-b** at 1500 K. The effect of varying molar-oxygen was explored as it emulates the experimental condition of exposing alloy to static air. The selective oxidation of Ti or Al in $Ti_2AlC+O_2$ can be written as:

$$\mathrm{Ti_2AlC} + 2\mathrm{y.O_2} \longrightarrow 2\mathrm{Ti_{2-2x}AlC} + 2\mathrm{y.TiO_2}, \text{ Eq. (1)}$$

and

$$4\mathrm{Ti_2AlC} + 3\mathrm{y.O_2} \longrightarrow 4\mathrm{Ti_2Al_{1-x}C} + 2\mathrm{y.Al_2O_3}. \text{ Eq. (2)}$$

The $TiO_2$ oxidizes to $CO_2$ on longer explore time to static air due to C diffusion, and reaction chain becomes:

$$\mathrm{Ti_2AlC} + \mathrm{O_2} \longrightarrow \mathrm{Al_2O_3} + \mathrm{TiO_2} + \mathrm{CO_2}, \text{ Eq. (3)}$$

For $Ti_2AlC+O_2$ oxidation, the O diffusion into $Ti_2AlC$ and Ti diffusion to the surface during the oxidation process works as the rate-limiting condition**.** Whereas reaction products during the selective oxidation of Al or Ti/Zr in $(Ti_{0.5}Zr_{0.5})_2AlC+O_2$ can be written as:

$$(\mathrm{Ti_{0.5}Zr_{0.5}})_2\mathrm{AlC} + 2\mathrm{y.O_2} \longrightarrow 2(\mathrm{Ti_{0.5}Zr_{0.5}})_{2-2x}\mathrm{AlC} + 2\mathrm{y.}(\mathrm{TiO_2} + \mathrm{ZrO_2}), \text{ Eq. (4)}$$

and

$$4(\mathrm{Ti_{0.5}Zr_{0.5}})_2\mathrm{AlC} + 3\mathrm{y.O_2} \longrightarrow 4\mathrm{Ti_2Al_{1-x}C} + 2\mathrm{y.Al_2O_3} + \mathrm{ZrAl_2}. \text{ Eq. (5)}$$

Here, $ZrO_2$ and $TiO_2$ oxidize to $CO_2$ on longer explore to static air because of C diffusion, and reaction chain can be written as:

$$(\mathrm{Ti_{0.5}Zr_{0.5}})_2\mathrm{AlC} + \mathrm{O_2} \longrightarrow \mathrm{TiO_2} + \mathrm{ZrO_2} + \mathrm{CO_2} + \mathrm{TiAl_2O_5}, \text{ Eq. (6)}$$

i.e., C from and Zr-C/Ti-C can diffuse through mixed Ti/Zr layers and oxidize. The diffusion of Ti/Zr to the surface and O into the $(Ti_{0.5}Zr_{0.5})_2AlC$ and corresponding reaction products work as the rate-limiting factor.

**Table 1**. Thermodynamically favorable phases and phase-fractions of (a) $Ti_2AlC$, and (b) $(Ti_{0.5}Zr_{0.5})_2AlC$ for varying oxygen content (0-11 moles) at 1500 K. In this work, we only considered the energies of elemental Al in fcc disorder phase. Liquid phase was out of scope due to lack of Al energies.

| **Stages** | **Mole-Oxygen** | **Reaction products and Phases-Fractions** | |
|---|---|---|---|
| | | **Phase** | **Phase-fraction** |
| **(a) $Ti_2AlC + O_2$** | | | |
| I. | 1 | $\mathbf{Al_2O_3}$ + $Ti_2AlC$ + $\mathbf{Ti_3AlC_2}$ + **TiO** | (0.17, 0.17, 0.33, 0.33) |
| | 1.75 | $Al_2O_3$ + $Ti_3AlC_2$ + **TiC** + TiO | (0.2, 0.17, 0.23, 0.4) |
| II. | 2.75 | $Al_2O_3$ + $\mathbf{Ti_2O_3}$ + TiC + TiO | (0.22, 0.11, 0.44, 0.22) |
| III. | 4 | $Al_2O_3$ + **C** + TiC + $Ti_2O_3$ | (0.21, 0.29, 0.14, 0.36) |
| | 4.75 | $Al_2O_3$ + C + $\mathbf{Ti_3O_5}$ + $Ti_2O_3$ | (0.22, 0.44, 0.22, 0.11) |
| IV. | 6.3 | $Al_2O_3$ + C + $\mathbf{CO_2}$ + $TiO_2$ | (0.14, 0.17, 0.11, 0.57) |
| | 7 | | (0.14, 0.07, 0.21, 0.57) |
| V. | 7.75 | $Al_2O_3$ + $CO_2$ + **O** + $TiO_2$ | (0.13, 0.27, 0.07, 0.53) |
| | 9 | | (0.10, 0.20, 0.30, 0.40) |
| | 11 | | (0.07, 0.14, 0.50, 0.29) |
| **(b) $(Ti_{0.5}Zr_{0.5})_2AlC+O_2$** | | | |
| I. | 1 | **Al + ZrC +** $\mathbf{ZrO_2}$ **+** $\mathbf{ZrAl_2}$ **+** $\mathbf{Ti_3AlC_2}$ | (0.20, 0.20, 0.30, 0.10,0.20) |
| II. | 1.35 | Al + ZrC + $ZrO_2$ + $\mathbf{Al_2O_3}$ + $Ti_3AlC_2$ | (0.33, 0.17, 0.33, 0.002,0.167) |
| | 2 | | (0.12, 0.18, 0.38, 0.13,0.19) |
| III. | 2.5 | ZrC + **TiC** + $ZrO_2$ + $Al_2O_3$ + $Ti_3AlC_2$ | (0.16, 0.08, 0.40, 0.20,0.16) |
| | 3.4 | | (0.01, 0.38, 0.40, 0.20,0.01) |
| IV. | 4 | **C** + TiC + $ZrO_2$ + $Al_2O_3$ + $\mathbf{TiAl_2O_5}$ | (0.10, 0.30, 0.40, 0.10,0.09) |
| V. | 4.75 | C + TiC + $\mathbf{Ti_2O_3}$ + $ZrO_2$ + $TiAl_2O_5$ | (0.26, 0.13, 0.03, 0.39,0.19) |
| VI. | 5.6 | C+ $\mathbf{CO_2}$ + $\mathbf{TiO_2}$ + $ZrO_2$ + $TiAl_2O_5$ | (0.31, 0.02, 0.17, 0.33,0.17) |
| | 7 | | (0.08, 0.25, 0.17, 0.33,0.17) |
| VII. | 9 | **O** + $CO_2$ + $TiO_2$ + $ZrO_2$ + $TiAl_2O_5$ | (0.33, 0.22, 0.11, 0.22,0.12) |
| | 11 | | (0.54, 0.15, 0.08, 0.15,0.08) |

The elemental chemical activity during oxidation in $Ti_2AlC+O_2$ are compared with disorder $(Ti_{0.5}Zr_{0.5})_2AlC+O_2$ in **Fig. 2** at 1500 K. In **Fig. 2a**, we show Ti/Al/C/O chemical potentials calculated at unknown molar fractions of reaction products by mixing of their $\Delta G_{form}$. We found that chemical activity of Al in **Fig. 2a** increases with increasing oxygen molar-fractions, which is directly related with the formation of protective oxide layer as found at all temperatures in **Fig. S4** and **Table 1a**. Based on chemical activities of Ti/Al/C/O, we identified two zones in **Fig. 2a** for $Ti_2AlC$ - (a) slow (I-III), and (b) sharp (IV-V) change in chemical potential. The sharp change in region IV-V occurs due to oxidation of C into gaseous $CO_2$.

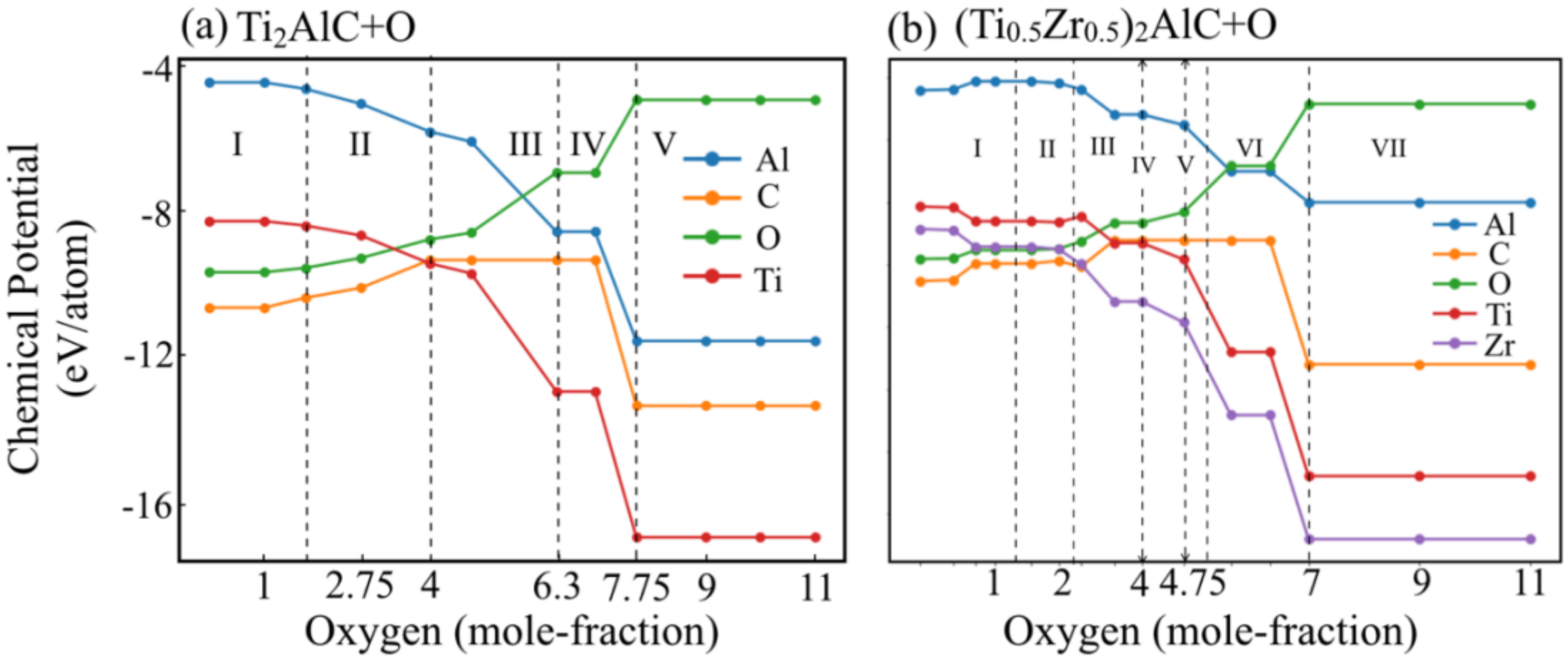

**Figure 2**: Change in elemental chemical activity of elements during oxidation of (a) $Ti_2AlC$, and (b) $(Ti_{0.5}Zr_{0.5})_2AlC$ at 1500 K as a function of molar oxygen.

The chemical activity of constituent elements of $(Ti_{0.5}Zr_{0.5})_2AlC+O_2$ was compared with $Ti_2AlC+O_2$ at 1500 K in **Fig. 2**. In **Fig. 2b**, Zr doping significantly changes the Al chemical activity in Ti2AlC that slows down the rate of protective $Al_2O_3$ formation. As shown in **Table 1b**, $Al_2O_3$ phase fraction increases from 1-4 moles oxygen then decreases and finally disappears, which is related to spinel $TiAl_2O_5$ that consumes all Al and leads to quick change in chemical activity **Fig. 2b**. We can see similar trends in disordered and ordered phases in **Fig. 2,** where Zr suppressed the Al activity. For moderate oxygen exposure in region I-V in **Fig. 2b**, very slow or no changes are observed for $(Ti_{0.5}Zr_{0.5})_2AlC$ as all the reaction products are still in solid phase. However, further increase in oxygen exposure, a jump in chemical activity of each element was observed that marks the beginning of the formation of gaseous phases such as C and $CO_2$ that marks the beginning of region V in **Fig. 2b**.

## Conclusions

We presented a detailed investigation of the oxidation mechanism for $(Ti_{0.5}Zr_{0.5})_2AlC$ MAX phase using high-throughput machine learning framework. We show that the M-site (at Ti) disordering of $Ti_2AlC$ by Zr has strong bearing on oxidation, which shows that Zr-doping significantly changes the chemical activity of $Ti_2AlC$ by slowing down the chemical activity of Al at higher oxygen exposure. The increased exposure of $(Ti_{0.5}Zr_{0.5})_2AlC$ to static air (oxygen) at higher temperature protects $Ti_3AlC_2$ up to moderate exposure to oxygen, however, increased oxygen exposure leads to the formation of spinel-phase ($TiAl_2O_5$) that may provide increased strengthening to $(Ti_{0.5}Zr_{0.5})_2AlC$. An ability to predict high temperature oxidation behavior makes our study useful to design MAX phases for future structural application.

## Acknowledgements

Project as supported through NSF Grant. No. 1729350. First-principles calculations were carried out at the Supercomputing Facility at Texas A&M University. DS acknowledges support from NSF through Grant No. 1545403.

## References.

1. M. Radovic, and M.W. Barsoum, *Am. Ceram. Soc. Bull.* **92**, 20-27 (2013).
2. Z.M. Sun, *Int. Mater. Rev.* **56**, 143-166 (2011).
3. P. Eklund, M. Beckers, U. Jansson, H. Högberg, L. Hultman, *Thin Solid Films* **518**, 1851-1878 (2010).
4. E. Hoffman et al., *Nucl. Eng. Des.* **244**, 17-24 (2012).
5. T.R. Allen, R.J.M. Konings, and A.T. Motta, in *Comprehensive Nuclear Materials*, Vol. 5 (ed. R. J. M. Konings) Ch. 5.03 (Elsevier, Amsterdam, 2012).
6. M.W. Barsoum, *MAX phases: properties of machinable ternary carbides and nitrides*. John Wiley & Sons, 2013.
7. M. Dahlqvist, B. Alling, and J. Rosén, *Phys. Rev. B* **81**, 220102 (2010).
8. X.H. Wang, and Y.C. Zhou, *Oxid. Met.* **59**, 303-320 (2003).
9. A.A. Balandin, *Nat. Mater.* **10**, 569–581 (2011).
10. R. Noriega et al., *Nat. Mater.* **12**, 1038 (2013).
11. C. Meneghini et al., *Phys. Rev. Lett.* **103**, 046403 (2009).
12. P. Singh, D. Sauceda, and R. Arroyave, *Acta Mater.* **184**, 50-58 (2019).
13. E.N. Hoffman, D.W.Vinson, R.L. Sindelar, D.J. Tallman, G. Kohse, and M.W. Barsoum, *Nuclear Engineering and Design* **244**, 17-24 (2012).

14. J.E. Saal, S. Kirklin, M. Aykol, B. Meredig, and C. Wolverton, *JOM* **65**, 1501-1509 (2013).
15. D. Sauceda, P. Singh, A.R. Falkowski, Y. Chen, T. Doung, G. Vazquez, M. Radovic, and R. Arroyave, npj Comp Mater **7**, 6 (2021).
16. A.R. Akbarzadeh, V. Ozolins, and C. Wolverton, *Adv. Mater.* ***19***, 3233-3239 (2007).
17. S. Kirklin, B. Meredig, and C. Wolverton, *Adv. Energy Mater*. **3**, 252-262 (2013).
18. A. van de Walle et al., *Calphad* **42**, 13-18 (2013).
19. G. Kresse, and J. Hafner, *Phys. Rev. B* **47**, 558-561 (1993).
20. G. Kresse, and D. Joubert, *Phys. Rev. B* **59**, 1758-1775 (1999).
21. H.J. Monkhorst, and J.D. Pack, *Phys. Rev. B* **13,** 5188-5192 (1976).
22. J.P. Perdew, K. Burke, and M. Ernzerhof, *Phys. Rev. Lett*. **77** 3865-3868 (1996).
23. C. Oses et al, *J. Chem. Inf. Model.* **58** (12), 2477-2490 (2018).